# Anapoles in Free-Standing III-V Nanodisks Enhancing Second-Harmonic Generation

*Maria Timofeeva[†,\*], Lukas Lang[†], Flavia Timpu[†], Claude Renaut[†], Alexei Bouravleuv[‡], Igor Shtrom[‡], George Cirlin[‡,§] and Rachel Grange[†]*

[†] ETH Zurich, Optical Nanomaterial Group, Institute for Quantum Electronics, Department of Physics, Auguste-Piccard Hof 1, 8093 Zurich, Switzerland

[‡] St. Petersburg Academic University, Ul. Khlopina 8/3, 194021 St. Petersburg, Russia

[§] ITMO University, Kronverkskiy 49, 197101 Saint Petersburg, Russia

*E-mail: mtimo@phys.ethz.ch

TABLE OF CONTENT GRAPHICS:

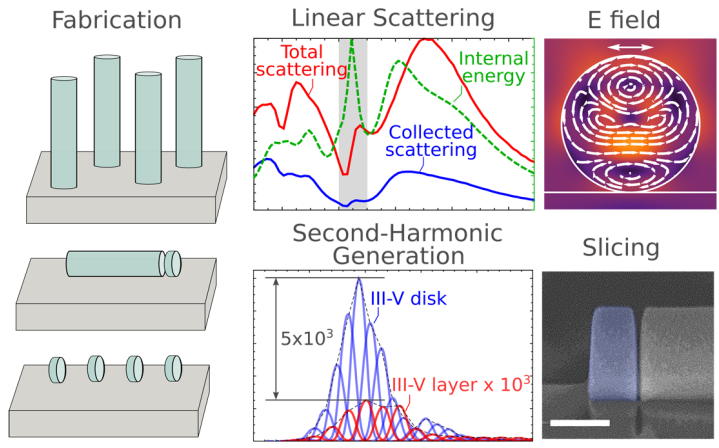

ABSTRACT:

Nonradiating electromagnetic configurations in nanostructures open new horizons for applications due to two essential features: lack of energy losses and invisibility to the propagating electromagnetic field. Such radiationless configurations form a basis for new types of nanophotonic devices, where a strong electromagnetic field confinement can be achieved together with lossless interactions between nearby components. In our work, we present a new design of free-standing disk nanoantennas with nonradiating current distributions for the optical near-infrared range. We show a novel approach to create nanoantennas by slicing III-V nanowires into standing disks using focused ion beam milling. We experimentally demonstrate the suppression of the far-field radiation and the associated strong enhancement of the second-harmonic generation from the disk nanoantennas. With a theoretical analysis of the electromagnetic field distribution using multipole expansions in both spherical and Cartesian coordinates, we confirm that the demonstrated nonradiating configurations are anapoles. We expect that the presented procedure to design and produce disk nanoantennas from nanowires become one of standard approaches to fabricate controlled chains of standing nanodisks with different designs and configurations. These chains can be an essential building blocks for new types of lasers and sensors with low power consumption.

The concept of radiationless electromagnetic configurations, which are dark and not detectable in the far-field, is still a significant theoretical and experimental challenge. These nonradiating configurations arise as a result of the destructive interference in the far-field between the radiation originating from two or more distinct current distributions inside the nanostructure[1]. In the last few years, significant results have been achieved in the theoretical analysis[2–5] and experimental demonstration[6–8] of anapole-type nonradiating current distributions. Initially, the term *anapole* was introduced in particle physics and then was used to describe dark matter in the universe[9]. In electrodynamics, first-order anapoles are nonradiating current distributions[10], where electric and toroidal dipole moments[11–13] destructively interfere in the far-field due to their identical radiation patterns[1,14], which results in the suppression of the far-field radiation. The excitation of anapole modes in nanostructures provides a key possibility to develop nonscattering metamaterials with strong electromagnetic field confinement inside[15,16]. This strong field confinement leads to a substantial increase of the nonlinear optical response, which can be employed both to detect anapoles and to design efficient nonlinear devices with low energy losses.

For the optical domain, the anapoles and toroidal modes were recently employed to enhance the third-order nonlinear optical effects in Ge disk nanoantennas[7,17] and were demonstrated in Si nanodisks[6]. In the microwave domain, dynamic anapoles were demonstrated experimentally and showed a great potential for anapole applications in new types of metamaterials[11,15,18]. The concept of anapoles was utilized to design a new type of laser based on InGaAs disk nanoantennas, which can be an ideal coherent light source for optics at the nanoscale due to the absence of losses and low power consumption[19]. The reduction of energy radiation and invisibility to the propagating electromagnetic field make anapoles an essential basis for a new type of devices with lossless interactions between nearby components. Anapoles address the fundamental tradeoff between the electromagnetic field confinement and energy dissipation, but a central challenge still is how to create nanostructures that utilize them effectively.

Our work is following the progressive research route[20,21] of nanoscale light manipulation with optically induced Mie resonances[15] in nanostructures from dielectric and semiconductor materials[24–26] with high refractive indexes as an alternative to plasmonic materials. This kind of nanostructures offer the possibilities to reduce dissipative losses significantly and perform high resonance enhancement of electric and magnetic fields. In addition to these advantages, semiconductor materials can also be electrically doped and used for subwavelength active devices[27–29]. Furthermore, compared with the widely used silicon nanostructures, III-V materials such as GaAs, AlGaAs, or InAs, have direct band gaps and non-centrosymmetric structure, which make them promising materials for nonlinear photonic devices.

Here, we develop a new concept of vertically free-standing disk nanoantennas fabricated from epitaxially grown III-V nanowires sliced with focused ion beam milling. This technique offers unique opportunities to fabricate high-quality structures with variable radii, longitudinal heterostructures from lattice-mismatched materials with different refractive indexes[30] and in crystal phases that are not available in bulk III-V materials[31–34].

We experimentally show the lack of radiation in the near-infrared domain and the strong second-harmonic generation (SHG) enhancement due to an efficient electromagnetic field confinement. The multipole expansion of the electromagnetic field, performed in both spherical[6,35] and Cartesian coordinates[36,37], shows that the fabricated standing disks have anapole-type nonradiating current distributions. The proposed novel geometry of vertically free-standing disks has a lower aspect ratio between diameter and height compared with previous horizontally lying disk designs[6,7] and results in significant decrease in the substrate influence. This configuration, with new types of material based on III-V nanowires[38], enables the use of any substrate regardless of the nanoantenna material and fabrication process. This is crucial for the development of nonlinear nanophotonic devices, and for compatibility with complementary metal-oxide-semiconductor (CMOS) technologies.

We fabricated vertically standing disk nanoantennas from core-shell GaAs/AlGaAs nanowires (see Methods), where AlGaAs and GaAs are non-centrosymmetric materials with strong second-order nonlinear optical properties[39–41]. Figure 1 presents the main steps of the fabrication process. We mechanically transferred nanowires, grown by molecular beam epitaxy[42], onto a glass substrate covered with indium tin oxide (ITO) (Figure 1a). Using focused ion beam (FIB) milling, we sliced them into standing disks with diameters of 440 nm and heights of 210 nm. Scanning electron microscopy (SEM) images of the disk during fabrication are shown in Figure 1b and 1c.

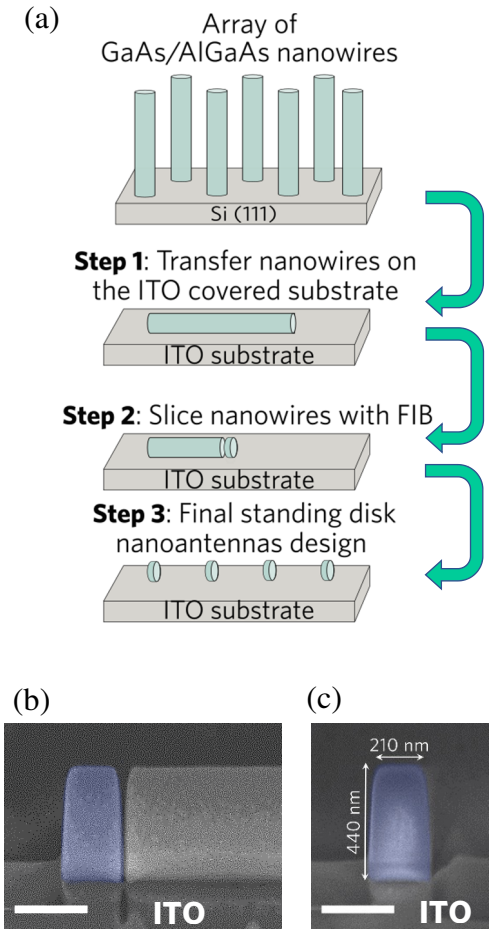

**Figure 1.** (a), Fabrication process flow using focused ion beam (FIB) milling for slicing the nanowires, placed on an indium tin oxide (ITO) substrate. (b), Scanning electron microscopy (SEM) image of one nanowire sliced by FIB into standing disks (highlighted in blue). (c), SEM image of the resulting single standing disk (blue) on an ITO substrate. Scale bars are 300 nm.

To demonstrate the ability of our design to support nonradiating current distributions, we developed a finite element method (FEM) model and studied two types of disk orientations on the glass substrate: vertically standing (Figure 2a) and horizontally lying (Figure 2b) disks. The FEM simulations were performed using the COMSOL Multiphysics® software. Our model considers a spherical system, where disk nanoantenna is placed in the center of the sphere, as shown on the Figure 2. The disk has as a cylindrical structure as highlighted in red in Figure 2 and it is placed on a glass substrate. The center of the disk is aligned with the center of the simulation sphere, where we perform our calculations. The background electric field $\vec{E}$ is a linearly polarized plane wave incident from the substrate side through the disk, as indicated by the wavevectors $\vec{k}$ in Figure 2 and is partially reflected at the substrate/air interface. The numerical aperture (NA) of the collection objective was accounted for by integrating the scattered signal over a limited solid angle, highlighted in blue in Figure 2.

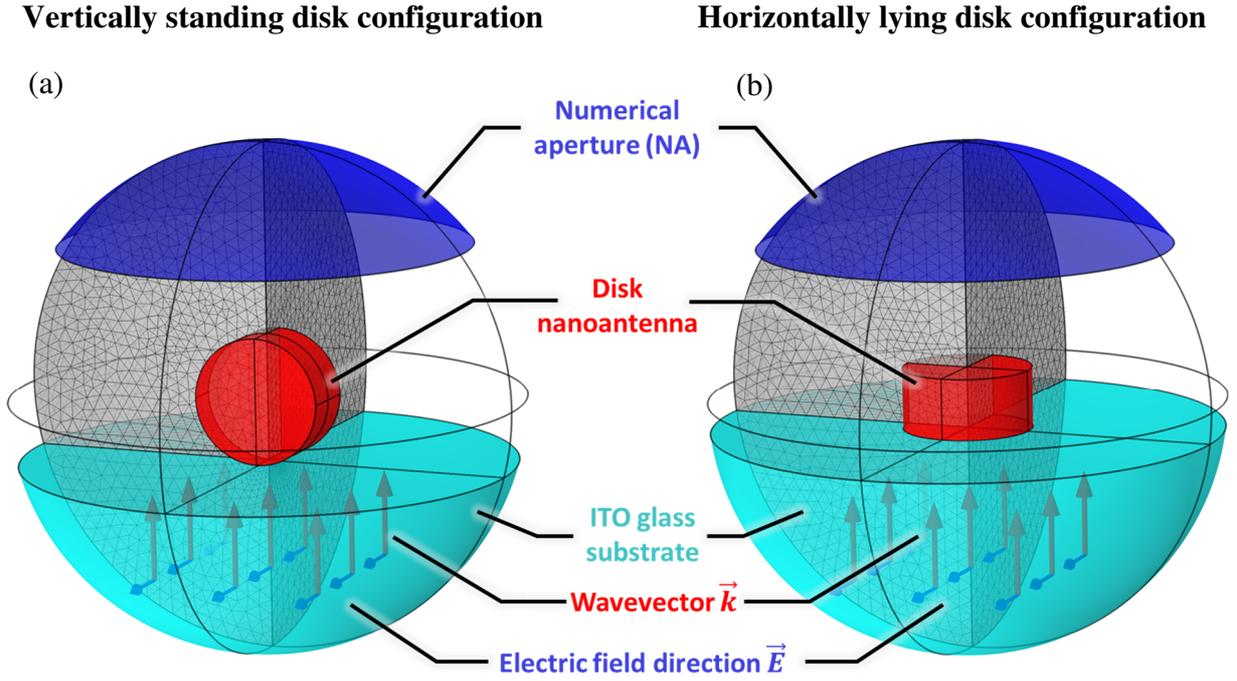

**Figure 2.** Numerical model schematic of the vertically standing configuration (a) and of the horizontally lying disk configuration (b).

In contrast to horizontally lying disks[6,17,43] (Figure 3a), the vertically standing configuration (Figure 3d, g) is sensitive to the polarization of the incident electric field $\vec{E}$. Therefore, we considered two electric field directions: perpendicular (Figure 3d) and parallel (Figure 3g) to the disk. Figures 3b, e, h presents the simulated normalized scattering cross sections for the corresponding orientations. These spectra were calculated both for a collection of the signal over the full sphere (red lines, Figures 3a, e, h) and via an objective with NA = 0.55 (blue lines, Figures 3b, e, h). All charts show the normalized internal electric energies, shown with green lines in Figures 3b, e, h. The results of the simulations demonstrate an important feature of the designed standing disks: the regions of the nonradiating current distributions are highly overlapping for both polarizations of the incoming electric field. Therefore, they will appear for unpolarized and arbitrarily polarized light as well.

The scattering cross section spectra from the vertically standing disks for both polarizations (red lines, Figures 3e, h) display sharp drops accompanied by the corresponding peaks of the internal electric energy (green lines, Figure 3e, h) between 800 and 900 nm (highlighted grey regions in Figure 3). This demonstrates an efficient electromagnetic field confinement with suppressed far-field scattering, indicating nonradiating current distributions[1,6]. We are not considering the other drops in the scattering spectra, as they are not accompanied by internal energy peaks. While the horizontally lying disk is the standard geometry for disk nanoantennas[6,17,43], the nonradiating configurations in such structures appear only for high aspect ratios, around 4:1 for Si[6] and Ge[7,17]. In our geometry, the nonradiating configurations are presented for an aspect ratio of 2:1, as shown on SEM image Figure 1c.

We calculated the electric field distributions inside the disks and we present the images for the corresponding results for the wavelengths of the internal energy peak in the region 800–900 nm in Figures 3c, f, i. The toroidal current distributions, shown in Figure 3f and i, indicate the presence of the toroidal mode inside the vertically standing nanoantennas[1].

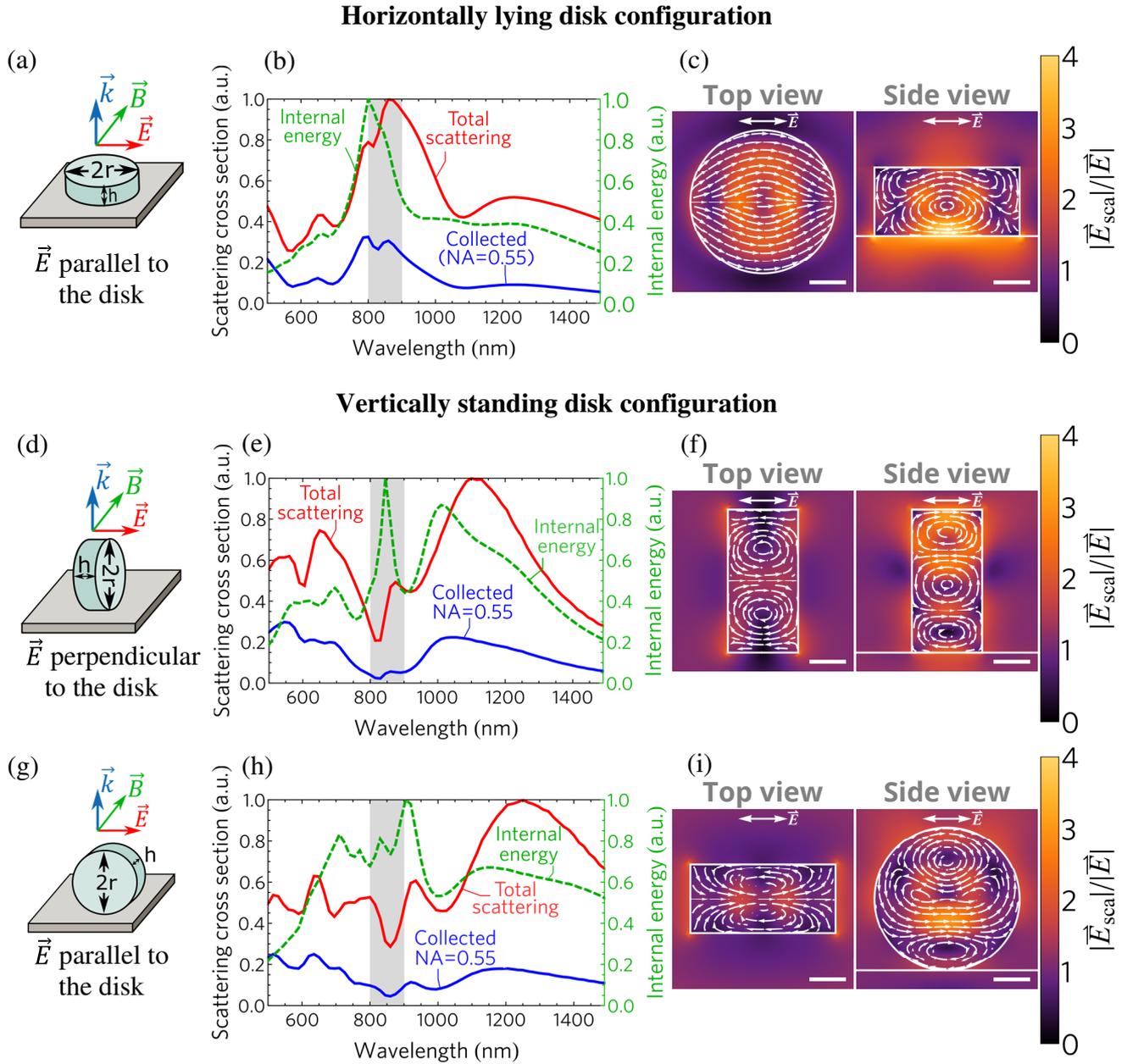

**Figure 3. Horizontally lying** disk configuration with (a), schematic, where vectors $\vec{k}$, $\vec{E}$ and $\vec{B}$ illustrate the directions of the wave vector, electric and magnetic fields, respectively, (b), simulated scattering cross section for full sphere collection of the signal (red lines), with collection via objective with numerical aperture (NA=0.55) and corresponding internal electric energy, and (c), top and side views for electric field distributions, the scale is 100 nm and $\vec{E}_{sca}$ is the scattered electric field. **Vertically standing disk configuration perpendicular to the incoming electric field** with the corresponding schematic (d), simulation results (e), and electric field distributions (f). **Vertically standing disk configuration parallel to the incoming electric field** with the corresponding schematic (g), simulation results (h), and electric field distributions (i). Grey area on (b), (e) and (h) highlights the region of interest.

To demonstrate experimentally the nonradiating current distribution in the fabricated vertically standing disks in the region 800–900 nm, we measured the linear and nonlinear optical responses. Figure 4a displays the scattering cross section from a single standing disk (Figure 1c) measured in a dark-field spectroscopy setup (see Supporting Information, section 3.1). Figure 4b shows the corresponding linear scattering cross section averaged over two polarizations calculated with FEM model. The simulated linear scattering cross section matches the experimental one, presented in Figures 4a, b, reproducing all major features.

The strong confinement of the electric field inside the vertically standing disks, demonstrated in numerical simulations (Figures 3f, i) for the region 800–900 nm, leads to a resonant enhancement of the SHG. We demonstrate this enhancement by measuring the SHG intensity spectra from the standing disk (blue lines, Figure 4c) and comparing them with the SHG spectra from an epitaxially grown layer of AlGaAs (red lines, Figure 4c). The total SHG enhancement factor of the standing disk is around $5 \times 10^3$ in comparison with the layer (see Supporting Information, Section 3.2). To analyze the SHG in the fabricated standing disks, we extended the FEM model with the corresponding SHG simulations (see Supporting Information, Section 1.1), and Figure 4d presents the results of these SHG simulations. The small red shifts between experimental and simulated SHG spectra could arise from deviations of shape and dimensions during the fabrication process in comparison with the perfect disk geometry in the FEM model.

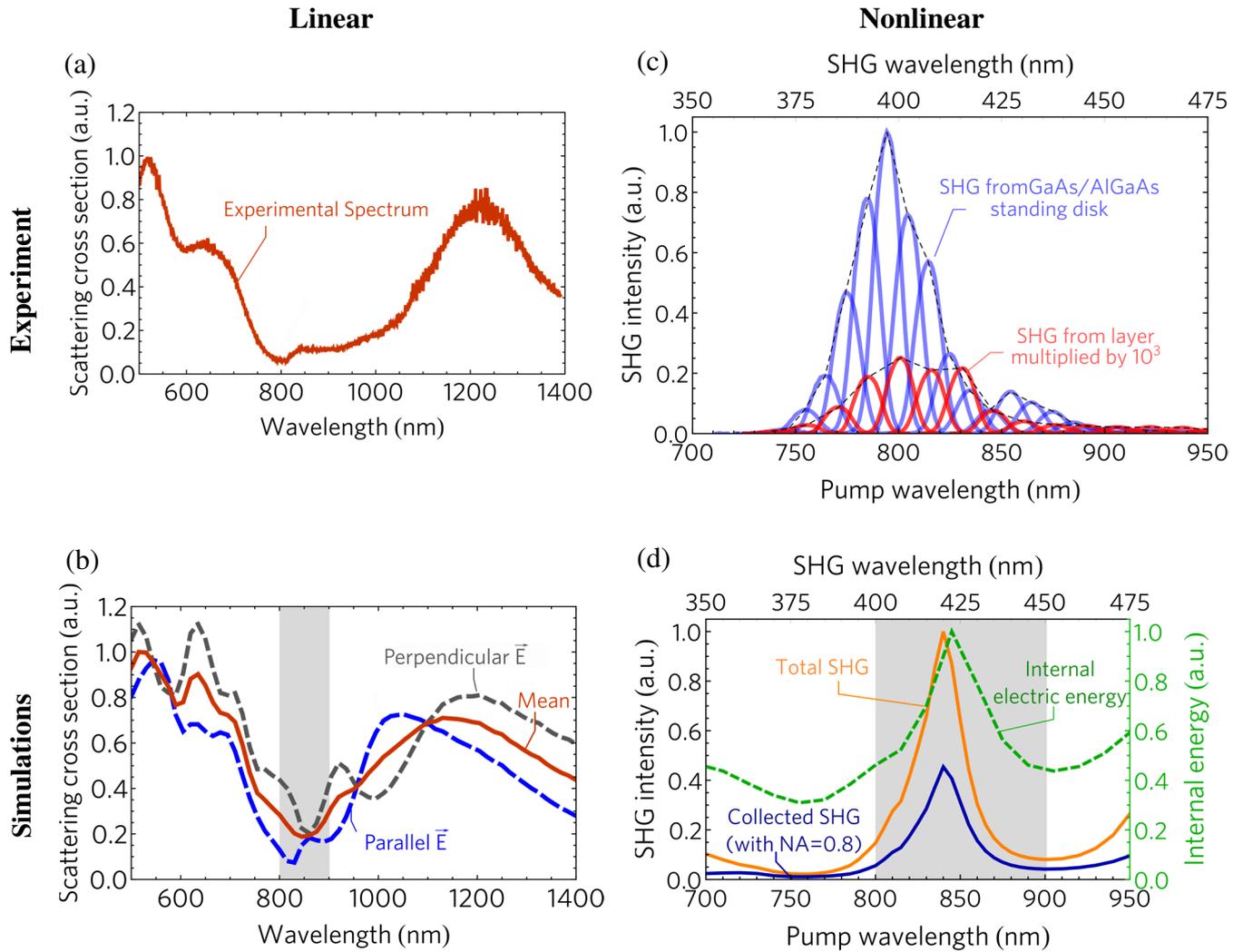

**Figure 4.** (a) Experimentally measured scattering cross section of an GaAs/AlGaAs disk nanoantenna. (b), Simulated scattering cross section spectra for both parallel (blue dashed line) and perpendicular (dashed grey line) polarization of the incoming electric field, and the average over both polarizations (solid red line). (c), Fundamental wavelength dependence of the SHG spectra of the GaAs/AlGaAs disk (blue lines) and epitaxial layer (red lines, signals were multiplied by $10^3$), the dashed grey lines are the envelope of the SHG spectra; SHG spectra were recorded by sweeping the excitation laser wavelength from 690 nm to 900 nm with steps of 10 nm. (d), Simulated SHG signal, for the full sphere collection (orange line), collection via objective with numerical aperture 0.8 (blue line), and the internal electric energy of the fundamental radiation (dashed green line). Grey area on (b) and (d) highlights the region of interest.

Simultaneous implementation of the multipole expansion of the electromagnetic field in spherical and Cartesian coordinates lets us unambiguously determine the nature of the nonradiating current distributions in the studied free-standing disks. In spherical coordinates, the far-field radiation can be easily decomposed up to higher-order multipoles. However, the toroidal multipoles share the same radiation pattern as the corresponding electric multipoles[1], which makes them indistinguishable in spherical coordinates, while the Cartesian expansion separates them (Supporting Information, Section 2.2).

For the spherical multipole expansion[35], we evaluated the contributions from electric (ED$_{sph}$) and magnetic (MD$_{sph}$) dipoles, electric (EQ$_{sph}$) and magnetic (MQ$_{sph}$) quadrupoles, and electric (EO$_{sph}$) and magnetic (MO$_{sph}$) octupoles (Figures 5a and c). The sum of these individual contributions, shown with red lines in Figures 5a, c, matches the calculated scattering cross section for the standing disk, shown with red lines in Figures 3b, c. As we can see from the spherical multipole expansions, the contributions from the spherical octupoles are negligible, which allows us to perform the Cartesian multipole expansion up to electric and magnetic quadrupoles terms only[36,37]. The corresponding expressions for the Cartesian multipoles contributions are presented in Table 1.

Table 1. Expressions for different Cartesian multipoles as integrals over the disk volume, where $J_{sca}(r)$ is the scattering current density (Supporting Information, section 2.2):

| Multipole | Symbol | Expression |
|---|---|---|
| Electric dipole | $p$ | $\frac{1}{i\omega}\int J_{sca}(r)d^3r$ |
| Magnetic dipole | $m$ | $\frac{1}{2c}\int r \times J_{sca}(r)d^3r$ |
| Toroidal dipole | $t$ | $\frac{1}{10c}\int r(r \cdot J_{sca}(r)) - 2r^2 J_{sca}(r)d^3r$ |
| Mean square radius | $\overline{\rho^2}$ | $\frac{1}{2c}\int r^2(r \times J_{sca})d^3r$ |
| Electric quadrupole | $Q_{\alpha\beta}$ | $\frac{1}{2i\omega}\int \left[r_\alpha J_{sca,\beta}(r) + r_\beta J_{sca,\alpha}(r) - \frac{2}{3}\delta_{\alpha,\beta}r J_{sca}(r)\right]d^3r$ |
| Magnetic quadrupole | $m_{i,j}$ | $\frac{1}{3c}\int \left[(r \times J_{sca}(r))_\alpha r_\beta + (r \times J_{sca}(r))_\beta r_\alpha\right]d^3r$ |

The total scattering power $I_{tot}$ and corresponding scattering cross section ($\sigma_{sc}$), where $\eta$ is the free-space impedance[36,37] is given by:

$$I_{tot} = \sigma_{sc}\frac{1}{2\eta}|E|^2 =$$

$$= \underbrace{\frac{2\omega^4}{6c^3}\left|\boldsymbol{p}+\frac{i\omega}{c}\boldsymbol{t}\right|^2}_{\text{TED}_{car}} + \underbrace{\frac{2\omega^4}{6c^3}|\boldsymbol{m}|^2}_{\text{MD}_{car}} - \underbrace{\frac{\omega^6}{15c^5}\text{Re}(\boldsymbol{m}^*\cdot\overline{\boldsymbol{\rho}^2})}_{\text{MR}_{car}} + \underbrace{\frac{\omega^6}{10c^5}|Q_{\alpha\beta}|^2}_{\text{EQ}_{car}} \quad (1)$$

$$+ \underbrace{\frac{\omega^6}{40c^5}|m_{\alpha\beta}|^2}_{\text{MQ}_{car}}$$

Here, the first term is the total electric dipole contribution (TED$_{car}$) and contains the electric (ED$_{car}$) and toroidal (TD$_{car}$) dipoles shown in purple and green lines in Figure 5e together with their interference that shown with blue line in Figure 5e. The TED$_{car}$ equals the electric dipole in the spherical coordinates (ED$_{sph}$)[2,35]. The second and third terms are the contribution from the magnetic dipole (MD$_{car}$) and the correction due to its interference with its mean square radius $\overline{\boldsymbol{\rho}^2}$ (MR$_{car}$) respectively. Their sum shown with dashed orange line in Figure 5b, neglecting the mean square radius term (eq. S10-11, Supporting Information), corresponds to the magnetic dipole (MD$_{sph}$) in spherical coordinates. The last two terms are the contributions from the electric (EQ$_{car}$) and magnetic (MQ$_{car}$) quadrupoles, that presented with solid and dashed blue lines in Figure 5d.

The spherical multipole expansion (Figures 5a and c) shows that the region with the drop of the total scattering cross section (800-900 nm) has a nonnegligible contribution only from the electric dipole ED$_{sph}$, while others are suppressed. The decomposition into Cartesian multipoles demonstrates high values of electric (ED$_{car}$) and toroidal dipoles (TD$_{car}$) in this region together with a strong destructive interference between them (Figure 5e) proportional to Im($\boldsymbol{p}^*\cdot\boldsymbol{t}$)[3,36,37]. The other multipoles, including the magnetic dipole (MD$_{car}$), in the Cartesian representation have no contributions in this region (Figures 5d,e). The multipole expansion demonstrates the destructive interference between electric and toroidal dipoles and simultaneous suppression of the higher-order multipoles. Therefore, our theoretical analysis shows the anapole nature of nonradiating current distributions in the designed and experimentally studied vertically standing disk nanoantennas.

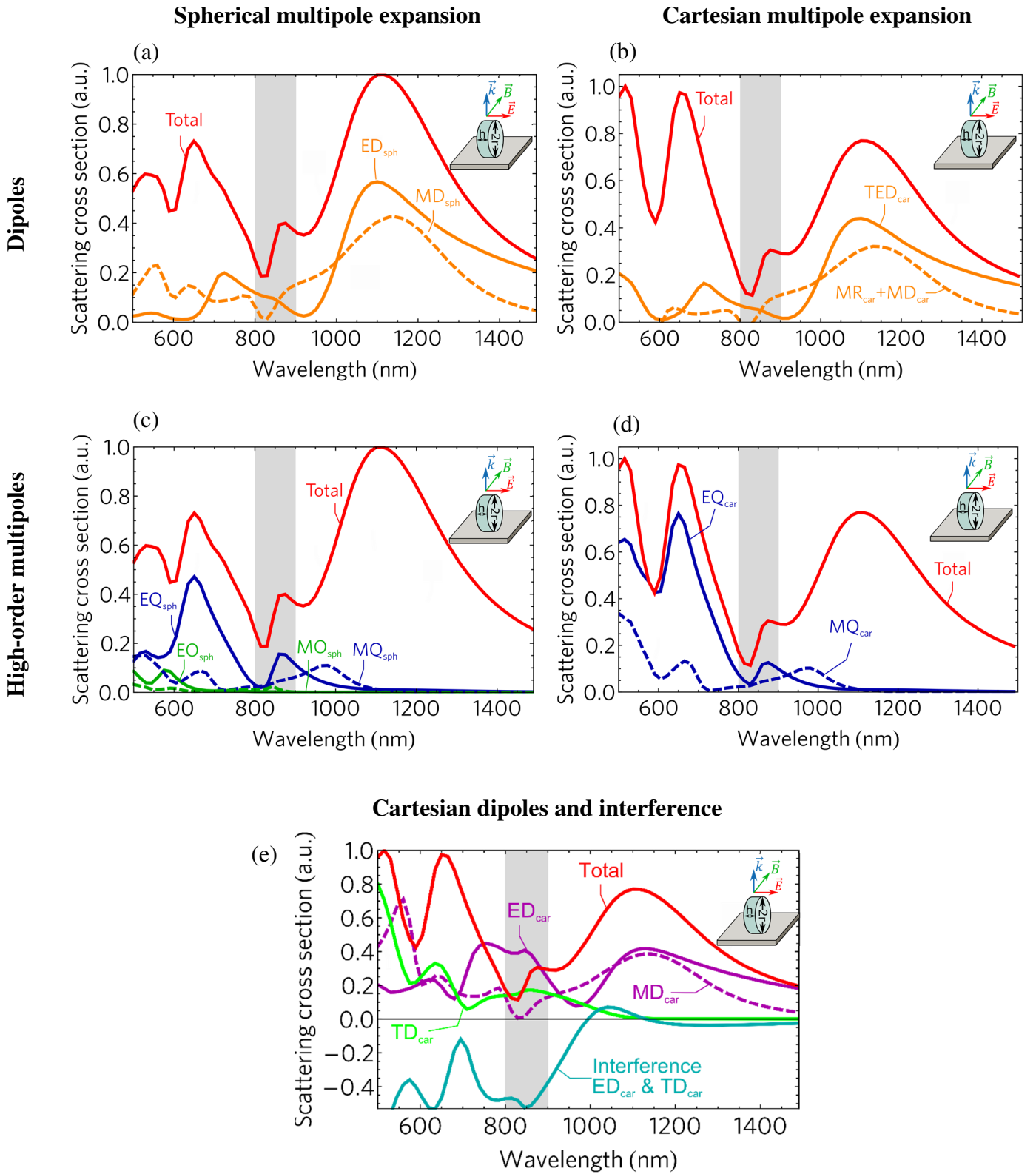

**Figure 5.** The geometry schematic for all charts is the same and presented on the insets. (a), Contributions from spherical electric ED$_{sph}$ and magnetic MD$_{sph}$ dipoles to the total scattering intensity (red line). (b), Contributions from Cartesian total electric dipole TED$_{car}$ and total magnetic dipole TMD$_{sph}$ (neglecting the mean square radius correction) to the total scattering intensity (red line). (c), Contributions from spherical electric EQ$_{sph}$ and magnetic MQ$_{sph}$ quadrupoles, electric EO$_{sph}$ and magnetic MO$_{sph}$ octupoles

to the total scattering intensity. (d), Contributions from Cartesian electric $EQ_{sph}$ and magnetic $MQ_{sph}$ quadrupoles. (e), Cartesian electric ($ED_{car}$) and magnetic ($MD_{car}$) dipoles, toroidal electric dipole ($TD_{car}$) and interference term of $ED_{car}$ & $TD_{car}$ (see equation (1)) to the total scattering cross section (red line). The negative value of the interference term (cyan line) arises since it is only part of the physical contribution of the total electric dipole $TED_{car}$.

In summary, we have presented a new design concept of nanoantennas based on free-standing disks, a novel procedure to fabricate them from III-V nanowires, and an end-to-end approach, from theory to experiment, to study nanostructures supporting nonradiating configurations in the optical range. The presented methods of slicing nanowires with FIB milling allows to create disk-based configurations on any substrate. This method significantly expands the possible application area of III-V disk nanoantennas. The approach of FIB nanowire slicing allows easily to create different designs of nanoantennas chains[25,44] without relying on the substrate material and to make them compatible with CMOS technologies. This method, presented in our work, is general and could be utilized for different types of nanowires and nanomaterials.

The presented end-to-end approach includes a numerical simulation with analysis of the multipole expansion of the electric field and experimental demonstration of nonradiating current distributions by linear scattering and SHG spectra measurements. We applied it for an experimental and theoretical demonstration of nonradiating anapoles in the geometry of vertically standing III-V nanodisks.

In the near future, we expect that the developed procedure of nanowires slicing can create a new branch of nanotechnology to fabricate controlled chains of standing nanodisks with different designs and configurations. Such type of nanoantennas' chains could be a core element for lasers and sensors[45,46] with low power consumption based on lossless nonlinear nanophotonic components supporting anapoles and compatible with CMOS technologies.

**Methods.** *Fabrication of GaAs/AlGaAs disks.* The vertically standing nanodisks were fabricated from GaAs/AlGaAs core-shell nanowires deposited on ITO glass substrate using FIB milling. The nanowires were grown by molecular beam epitaxy on Si (111) substrates, using Au colloid droplets as a catalyst material[42]. As the first step, the GaAs core of nanowires were grown, then in situ the $Al_{0.2}Ga_{0.8}As$ shells were grown to passivate the surface states[47]. Then nanowires were transferred on the ITO covered glass substrate. After that, they were patterned using a cross beam electron and ion microscope (Zeiss). This system combines FIB and field emission SEM. The FIB column uses liquid metal $Ga^+$ as ion source. Using the special software, nanowires could be patterned in different nanoantenna configurations.

*Far-field linear spectroscopy.* Characterization of linear scattering cross sections were performed in a dark-field spectroscopy setup[48] (Supporting Information, section 3.1). The nanoantenna sample was irradiated with a halogen lamp source. The scattered light was collected in a transmission geometry using a 50x objective with a numerical aperture NA = 0.55.

*Nonlinear optical characterization.* Nonlinear optical measurements of the SHG spectra were performed with a nonscanning transmission optical microscope (Supporting Information, Figure S2). For the excitation of the samples, we used a tunable Ti:Sapphire laser (in the range 690−900 nm with steps of 10 nm). The laser light was focused with a 10x objective on the ITO covered glass with isolated disk nanoantennas. The signal was then collected with a 100x objective (NA= 0.8) and focused onto a scientific CMOS camera and on the spectrometer. The polarization of the excitation laser was fixed the same for all steps. As a reference sample for the determination of the enhancement factor, we measured the SHG spectra from an epitaxially grown layer of AlGaAs with a thickness of 800 nm. The corresponding spectra from the layer and the disks were normalized by the effective volume of the measured structures and the differences in the setup parameters for reflection and transmission measurements (Supporting Information, section 3.2).

*Numerical FEM simulations.* The numerical simulations were performed using the COMSOL Multiphysics® software package. For the numerical simulations of the scattering cross section from the disk structures we developed a model to calculate the linear scattering cross section from the disk nanoantennas and the spectra of SHG for different geometries. This model considers the numerical aperture of the collection objectives.

*Multipole expansions.* Multipole expansions were performed in both spherical and Cartesian coordinate systems. We analyze the contributions to the total scattering cross section from the individual multipoles up to octupoles terms in spherical coordinates and up to order $\frac{1}{c^5}$ for Cartesian ones[37]. The spherical and Cartesian multipole expansions were implemented in the developed COMSOL model (Supporting Information, section 2).

ASSOSIATED CONTENT

**Supporting Information**

More details about experimental methods and theoretical analysis with multipole expansions in spherical and Cartesian coordinates are presented in Supporting Information. Numerical simulations of SHG; numerical COMSOL models for multipole expansions in spherical and Cartesian coordinates; experimental setups for linear and nonlinear characterizations.

These material is available free of charge via Internet at http://pubs.acs.org.

AUTHOR INFORMATION

**Corresponding Author**

*E-mail: mtimo@phys.ethz.ch

*E-mail: grange@phys.ethz.ch

**Author Contributions**

The manuscript was written through contributions of all authors. All authors have given approval to the final version of the manuscript.

R.G. designed the experiment. M.T. did the nanoantennas design, second-harmonic generation experiments, data analysis, theoretical modelling and manuscript writing. L.L. performed the second-harmonic generation measurements, numerical simulations and theoretical analysis. F.T. performed the linear optical characterization and helped with numerical simulations. C.R. did the focused ion beam milling of nanowires and characterization with scanning electron microscopy. A.B., Ig.S. and G.C. fabricated the nanowires with molecular beam epitaxy. All authors have given approval to the final version.

ACKNOWLEDGMENT

The authors thank the Scientific Centre for Optical and Electron Microscopy (ScopeM) of ETH Zurich, the financial supports from the Swiss National Science Foundation (SNF) (150609), Bridge project No 173829, and the ERC starting grant Chi2-Nano-Oxide. The nanowire samples were grown under the support of Russian Science Foundation (RSF) Project No 14-12-00393.

ABBREVIATIONS

Si, silicon; Ge, germanium; InGaAs, indium gallium arsenide; GaAs, gallium arsenide, AlGaAs, aluminium gallium arsenide, InAs, indium arsenide; CMOS, complementary metal-oxide-semiconductor; SHG, second-harmonic generation; ITO, indium tin oxide; FIB, focused ion beam; SEM, scanning electron microscopy, FEM, finite element model; NA, numerical aperture; $ED_{sph}$, spherical electric dipole; $MD_{sph}$, spherical magnetic dipole; $EQ_{sph}$, spherical electric quadrupole; $MQ_{sph}$, spherical magnetic quadrupole; $EO_{sph}$, spherical magnetic octupole; $MO_{sph}$, spherical magnetic octupole; $TED_{car}$, cartesian total electric dipole contribution; $ED_{car}$, Cartesian electric dipole; $TD_{car}$ toroidal dipole; $MD_{car}$, Cartesian magnetic dipole; $MR_{car}$, mean square radius; $EQ_{car}$, Cartesian electric quadrupole; $MQ_{car}$, Cartesian magnetic quadrupole.

REFERENCES

(1) Papasimakis, N.; Fedotov, V. A.; Savinov, V.; Raybould, T. A.; Zheludev, N. I. Electromagnetic Toroidal Excitations in Matter and Free Space. *Nat. Mater.* **2016**, *15* (3), 263–271.

(2) Evlyukhin, A. B.; Fischer, T.; Reinhardt, C.; Chichkov, B. N. Optical Theorem and Multipole Scattering of Light by Arbitrarily Shaped Nanoparticles. *Phys. Rev. B* **2016**, *94* (20), 1–7.

(3) Raybould, T.; Fedotov, V. A.; Papasimakis, N.; Youngs, I.; Zheludev, N. I. Exciting Dynamic Anapoles with Electromagnetic Doughnut Pulses. *Appl. Phys. Lett.* **2017**, *111* (8).

(4) Luk'yanchuk, B.; Paniagua-Domínguez, R.; Kuznetsov, A. I.; Miroshnichenko, A. E.; Kivshar, Y. S. Hybrid Anapole Modes of High-Index Dielectric Nanoparticles. *Phys. Rev. A* **2017**, *95* (6), 1–8.

(5) Luk'yanchuk, B.; Paniagua-Domínguez, R.; Kuznetsov, A. I.; Miroshnichenko, A. E.; Kivshar, Y. S. Suppression of Scattering for Small Dielectric Particles: Anapole Mode and Invisibility. *Philos. Trans. R. Soc. A Math. Phys. Eng. Sci.* **2017**, *375* (2090), 20160069.

(6) Miroshnichenko, A. E.; Evlyukhin, A. B.; Yu, Y. F.; Bakker, R. M.; Chipouline, A.; Kuznetsov, A. I.; Luk'yanchuk, B.; Chichkov, B. N.; Kivshar, Y. S. Nonradiating Anapole Modes in Dielectric Nanoparticles. *Nat. Commun.* **2015**, *6*, 1–8.

(7) Grinblat, G.; Li, Y.; Nielsen, M. P.; Oulton, R. F.; Maier, S. A. Enhanced Third Harmonic Generation in Single Germanium Nanodisks Excited at the Anapole Mode. *Nano Lett.* **2016**, *16*, 4635–4640.

(8) Grinblat, G.; Li, Y.; Nielsen, M. P.; Oulton, R. F.; Maier, S. A. Enhanced Third Harmonic Generation in Single Germanium Nanodisks Excited at the Anapole Mode. *Nano Lett.* **2016**, *16* (7), 4635–4640.

(9) Zel'dovich, I. B. Electromagnetic Interaction with Parity Violation. *J. Exptl. Theor. Phys.* **1957**,

*33*, 1531–1533.

(10) Devaney, A. J.; Wolf, E. Radiating and Nonradiating Classical Current Distributions and the Fields They Generate. *Phys. Rev. D* **1973**, *8* (4), 1044–1047.

(11) Kaelberer, T.; Fedotov, V. A.; Papasimakis, N.; Tsai, D. P.; Zheludev, N. I. Toroidal Dipolar Response in a Metamaterial. *Science (80-. ).* **2010**, *330* (6010), 1510–1512.

(12) Savinov, V.; Fedotov, V. A.; Zheludev, N. I. Toroidal Dipolar Excitation and Macroscopic Electromagnetic Properties of Metamaterials. *Phys. Rev. B - Condens. Matter Mater. Phys.* **2014**, *89* (20).

(13) Basharin, A. A.; Kafesaki, M.; Economou, E. N.; Soukoulis, C. M.; Fedotov, V. A.; Savinov, V.; Zheludev, N. I. Dielectric Metamaterials with Toroidal Dipolar Response. *Phys. Rev. X* **2015**, *5* (1), 1–11.

(14) Afanasiev, G. N.; Stepanovsky, Y. P. The Electromagnetic Field of Elementary Time-Dependent Toroidal Sources. *J. Phys. A. Math. Gen.* **1995**, *28* (16), 4565–4580.

(15) Fedotov, V. A.; Rogacheva, A. V.; Savinov, V.; Tsai, D. P.; Zheludev, N. I. Resonant Transparency and Non-Trivial Non-Radiating Excitations in Toroidal Metamaterials. *Sci. Rep.* **2013**, *3*, 1–5.

(16) Gupta, M.; Savinov, V.; Xu, N.; Cong, L.; Dayal, G.; Wang, S.; Zhang, W.; Zheludev, N. I.; Singh, R. Sharp Toroidal Resonances in Planar Terahertz Metasurfaces. *Adv. Mater.* **2016**, 8206–8211.

(17) Grinblat, G.; Li, Y.; Nielsen, M. P.; Oulton, R. F.; Maier, S. A. Efficient Third Harmonic Generation and Nonlinear Subwavelength Imaging at a Higher-Order Anapole Mode in a Single Germanium Nanodisk. *ACS Nano* **2017**, *11* (1), 953–960.

(18) Marinov, K.; Boardman, A. D.; Fedotov, V. A.; Zheludev, N. Toroidal Metamaterial. *New J. Phys.* **2007**, *9*.

(19) Totero Gongora, J. S.; Miroshnichenko, A. E.; Kivshar, Y. S.; Fratalocchi, A. Anapole Nanolasers

for Mode-Locking and Ultrafast Pulse Generation. *Nat. Commun.* **2017**, *8*, 1–9.

(20) Staude, I.; Schilling, J. Metamaterial-Inspired Silicon Nanophotonics. *Nat. Photonics* **2017**, *11* (5), 274–284.

(21) Jahani, S.; Jacob, Z. All-Dielectric Metamaterials. *Nat. Publ. Gr.* **2016**, *11* (1), 23–36.

(22) Zhao, Q.; Zhou, J.; Zhang, F.; Lippens, D. Mie Resonance-Based Dielectric Metamaterials. *Mater. Today* **2009**, *12* (12), 60–69.

(23) Kivshar, Y.; Miroshnichenko, A. E. Meta- Optics with Mie Resonances. *Opt. Photonics News* **2017**, 24–31.

(24) Evlyukhin, A. B.; Reinhardt, C.; Seidel, A.; Luk'Yanchuk, B. S.; Chichkov, B. N. Optical Response Features of Si-Nanoparticle Arrays. *Phys. Rev. B - Condens. Matter Mater. Phys.* **2010**, *82* (4), 1–12.

(25) Krasnok, A. E.; Miroshnichenko, A. E.; Belov, P. a; Kivshar, Y. S. All-Dielectric Optical Nanoantennas. *Opt. Express* **2012**, *20* (18), 20599–20604.

(26) Valentine, J.; Li, J.; Zentgraf, T.; Bartal, G.; Zhang, X. An Optical Cloak Made of Dielectrics. *Nat. Mater.* **2009**, *8* (7), 568–571.

(27) Cao, L.; White, J. S.; Park, J.-S.; Schuller, J. A.; Clemens, B. M.; Brongersma, M. L. Engineering Light Absorption in Semiconductor Nanowire Devices. *Nat. Mater.* **2009**, *8* (8), 643–647.

(28) Garín, M.; Fenollosa, R.; Alcubilla, R.; Shi, L.; Marsal, L. F.; Meseguer, F. All-Silicon Spherical-Mie-Resonator Photodiode with Spectral Response in the Infrared Region. *Nat. Commun.* **2014**, *5*, 3440.

(29) Grzela, G.; Paniagua-Domínguez, R.; Barten, T.; Fontana, Y.; Sánchez-Gil, J. A.; Gómez Rivas, J. Nanowire Antenna Emission. *Nano Lett.* **2012**, *12* (11), 5481–5486.

(30) Assali, S.; Van Dam, D.; Haverkort, J. E. M.; Bakkers, E. P. A. M. High Refractive Index in

Wurtzite GaP Measured from Fabry-Pérot Resonances. *Appl. Phys. Lett.* **2016**, *108* (17), 1–4.

(31) Dick, K. A.; Thelander, C.; Samuelson, L.; Caroff, P. Crystal Phase Engineering in Single InAs Nanowires. *Nano Lett.* **2010**, *10* (9), 3494–3499.

(32) Lehmann, S.; Jacobsson, D.; Dick, K. A. Crystal Phase Control in GaAs Nanowires : Opposing Trends in the Ga- and As-Limited Growth Regimes. *Nanotechnology* **2015**, *26* (30), 301001.

(33) Jacobsson, D.; Panciera, F.; Tersoff, J.; Reuter, M. C.; Lehmann, S.; Hofmann, S.; Dick, K. A.; Ross, F. M. Interface Dynamics and Crystal Phase Switching in GaAs Nanowires. *Nature* **2016**, *531* (7594), 317–322.

(34) Gudiksen, M. S.; Lauhon, L. J.; Wang, J.; Smith, D. C.; Lieber, C. M. Growth of Nanowire Superlattice Structures for Nanoscale Photonics and Electronics. *Lett. to Nat.* **2002**, *415*, 617–620.

(35) Grahn, P.; Shevchenko, A.; Kaivola, M. Electromagnetic Multipole Theory for Optical Nanomaterials. *New J. Phys.* **2012**, *14*, 93033.

(36) Radescu, E. E.; Vaman, G. Exact Calculation of the Angular Momentum Loss, Recoil Force, and Radiation Intensity for an Arbitrary Source in Terms of Electric, Magnetic, and Toroid Multipoles. *Phys. Rev. E - Stat. Physics, Plasmas, Fluids, Relat. Interdiscip. Top.* **2002**, *65* (4), 046609 (47 p).

(37) Radescu, E.; Vaman, G. Cartesian Multipole Expansions and Tensorial Identities. *Prog. Electromagn. Res. B* **2012**, *36*, 89–111.

(38) Vukajlovic-Plestina, J.; Kim, W.; Dubrovski, V. G.; Tütüncüoğlu, G.; Lagier, M.; Potts, H.; Friedl, M.; FontcubertaMorral, A. Engineering the Size Distributions of Ordered GaAs Nanowires on Silicon. *Nano Lett.* **2017**, *17* (7), 4101–4108.

(39) Timofeeva, M.; Bouravleuv, A.; Cirlin, G.; Shtrom, I.; Soshnikov, I.; Reig, M.; Sergeyev, A.; Grange, R. Polar Second-Harmonic Imaging to Resolve Pure and Mixed Crystal Phases Along GaAs Nanowires. *Nano Lett.* **2016**, *10*, 6290–6297.

(40) Liu, S.; Sinclair, M. B.; Saravi, S.; Keeler, G. A.; Yang, Y.; Reno, J.; Peake, G. M.; Setzpfandt,

F.; Staude, I.; Pertsch, T.; et al. Resonantly Enhanced Second-Harmonic Generation Using III-V Semiconductor All-Dielectric Metasurfaces. *Nano Lett.* **2016**, *16* (9), 5426–5432.

(41) Grange, R.; Bro, G.; Kiometzis, M.; Sergeyev, A.; Richter, J.; Leiterer, C.; Fritzsche, W.; Gutsche, C.; Lysov, A.; Prost, W.; et al. Far-Field Imaging for Direct Visualization of Light Interferences in GaAs Nanowires. *Nano Lett.* **2012**, *12* (10), 5412–5417.

(42) Bouravleuv, A. D.; Ilkiv, I.; Reznik, R.; Kotlyar, K.; Soshnikov, I. P.; Cirlin, G. E.; Brunkov, P.; Kirilenko, D.; Bondarenko, L.; Nepomnyaschiy, A.; et al. New Method for MBE Growth of GaAs Nanowires on Silicon Using Colloidal Au Nanoparticles. *Nanotechnology* **2017**, 1–17.

(43) Wang, R.; Dal Negro, L. Engineering Non-Radiative Anapole Modes for Broadband Absorption Enhancement of Light. *Opt. Express* **2016**, *24* (17), 19048.

(44) Li, S. V.; Krasnok, A. E.; Lepeshov, S.; Savelev, R. S.; Baranov, D. G.; Alu, A. All-Optical Switching and Unidirectional Plasmon Launching with Electron-Hole Plasma Driven Silicon Nanoantennas. *Phys. Rev. Appl.* **2017**, *9* (1), 14015.

(45) Patolsky, F.; Zheng, G.; Lieber, C. M. Nanowire-Based Biosensors. *Anal. Chem.* **2006**, *78* (13), 4261–4269.

(46) Patolsky, F.; Lieber, C. M. Nanowire Nanosensors. *Mater. Today* **2005**, *8* (4), 20–28.

(47) Demichel, O.; Heiss, M.; Bleuse, J.; Mariette, H.; Fontcuberta Morral, I. A. Impact of Surfaces on the Optical Properties of GaAs Nanowires. *Appl. Phys. Lett.* **2010**, *97* (20), 1–4.

(48) Timpu, F.; Hendricks, N. R.; Petrov, M.; Ni, S.; Renaut, C.; Wolf, H.; Isa, L.; Kivshar, Y.; Grange, R. Enhanced Second-Harmonic Generation from Sequential Capillarity- Assisted Particle Assembly of Hybrid Nanodimers. *Nano Lett.* **2017**, *17* (9), 5381–5388.